\documentclass[review]{elsarticle}

\newcommand{\F}{\mathbb{F}}

\usepackage{mathrsfs}
\usepackage{amssymb}
\usepackage{amsmath}
\usepackage{amsthm}
\usepackage{latexsym}
\usepackage{indentfirst}
\usepackage{enumitem}
\usepackage{anysize}
\usepackage{bbm}

\usepackage{lineno,hyperref}
\modulolinenumbers[5]

\journal{Journal of \LaTeX\ Templates}
\begin{document}

\begin{frontmatter}

\title{Trace Codes with Few Weights over $\F_p+u\F_p$ }



\author[mymainaddress]{Minjia Shi$^{*}$\fnref{myfootnote}}
\ead{smjwcl.good@163.com}

\author{Yan Liu$^{a}$}

\fntext[myfootnote]{The author is supported by NNSF of China (61672036),
Technology Foundation for Selected Overseas Chinese Scholar, Ministry of Personnel of China (05015133), the Open Research Fund of National Mobile Communications Research Laboratory, Southeast University (2015D11) and Key projects of support program for outstanding young talents in Colleges and Universities (gxyqZD2016008).}
\ead{liuyan2612@126.com}

\author[mysecondaryaddress]{Patrick Sol\'e}
\cortext[mycorrespondingauthor]{Corresponding author}
\ead{sole@enst.fr}

\address[mymainaddress]{Key Laboratory of Intelligent Computing $\&$ Signal Processing,
Ministry of Education, Anhui University No. 3 Feixi Road, Hefei Anhui Province 230039, P. R. China, National Mobile Communications Research Laboratory, Southeast University and School of Mathematical Sciences of Anhui University,
Anhui, 230601, P. R. China}
\address[mysecondaryaddress]{CNRS/LAGA, University Paris 8, 93 526 Saint-Denis, France}

%
%
%
%
%
%
%
\begin{abstract}
We construct an infinite family of two-Lee-weight and three-Lee-weight codes over the chain ring $\mathbb{F}_p+u\mathbb{F}_p.$
They have the algebraic structure of abelian codes. Their Lee weight distribution is computed by using Gauss sums.
 Then by using a linear Gray map, we obtain an infinite family of abelian codes with few weights over $\F_p$. In particular, we obtain an infinite family of two-weight codes which meets the Griesmer bound with equality.
 Finally, an application to secret sharing schemes is given.
\end{abstract}

\end{frontmatter}

\linenumbers

\section{Introduction}
Linear codes with few weights have applications in secret sharing \cite{DY2},
combinatorial designs, graph theory \cite{BGH}, association schemes, and difference sets \cite{CG,CW} as well, in addition to their applications in data storage systems. Hence, a code with few weights is a very interesting topic, and has been
investigated in \cite{D,DLLZ, DY,HY}  and many other papers.

A classical construction of codes over finite fields called {\bf trace codes} is as follows.
Let $$D=\{d_1,d_2,\dots,d_n\}\subseteq \F_{q^m}^*,$$ with $m$ a positive integer. A $q$-ary linear code of
length $n$ is defined by $$C_D=\{(tr(xd_1),tr(xd_2),\dots,tr(xd_n)):x\in \F_{q^m} \},$$ where $tr()$ is the so-called absolute trace function from $\F_{q^m}$ down to $\F_{q}.$
Here $D$ is called the {\bf defining set} of $C_D$. If the set $D$ is well chosen,
$C_D$ may have good parameters.
Many known codes \cite{HY1, HY} can be produced by selecting suitably the defining set $D.$

In a series of papers \cite{SLS1,SLS2,SWLP}, we have extended the notion of trace codes from fields to rings as follows. If $R$ is a finite ring, and $R_m$ an extension of $R$ of degree $m,$ we construct
a trace code with a defining set $L=\{d_1,d_2,\dots,d_{n'}\}\subseteq R_m^*,$ the group of units of $R_m$ by the formula
$$C_L=\{(Tr(xd_1),Tr(xd_2),\dots,Tr(xd_{n'})):x\in R_m \}=\{(Tr(xd))_{d \in L}: x\in R_m\},$$ where $Tr()$ is a linear function from $R_m$ down to $R.$
By varying $L$ and $R,$ various codes can be constructed. The obvious choice for $L$ is $R_m^*$ itself. In the present paper, we will consider a subgroup of index $2,$ leading
naturally to use quadratic Gauss  sums and Gauss periods. Note that if $L$ is a cyclic subgroup of $R_m^*$, the trace code will be cyclic, while if $L$ is an abelian group, the trace code will be merely abelian. In the present paper, we use a defining set inspired from that of the trace codes in \cite{HY1, HY} over the ring $R=\F_p+u\F_p.$ We can summarize this paragraph as shown below.

\begin{itemize}
\item[\cite{SLS1}] $L=R_m^*,$ $R=\F_2+u\F_2$,
\item[\cite{SLS2}] $L=R_m^*,$ $R=\F_2+u\F_2+v\F_2+uv\F_2$,
\item[\cite{SWLP}] $[R_m^*:L]=2,$ $R=\F_p+u\F_p$.
\item[Here] $L=D+u \F_{p^m},$ $R=\F_p+u\F_p$.
\end{itemize}

 Furthermore,
an application to secret sharing schemes is sketched out.

The rest of this paper is organized as follows. The next section describes the basic notations and known facts, which will be needed in Section 4. Section 3 shows that the codes are abelian. Section 4 gives the main results in this paper, the Lee weight distribution of our codes. Furthermore, the Gray images of two-Lee-weight codes meet the Griesmer bound with equality. Our codes have applications in secret sharing schemes in section 5. We will sum up all we have done throughout this paper in section 6, and make some conjectures for future research.

\section{Preliminaries}
\subsection{Rings}

Throughout this paper, let $p$ denote an odd prime. Denote by $R=\F_p+u\F_p$ the finite chain ring with $p^2$ elements, where $u^2=0.$ Given a positive integer $m$, we can construct the ring extension ${\mathcal{R}}=\F_{p^m}+u\F_{p^m}.$ For convenience, the set $M$ denotes its maximal ideal, i.e., $M=\{bu:b\in \F_{p^m}\}$. The residue field $\mathcal{R}/M$ is isomorphic to $\F_{p^m}$.
The group of units in ${\mathcal{R}},$ denoted by  ${\mathcal{R}}^*,$ is $\{a+bu:a\in \F_{p^m}^{*},b\in \F_{p^m}\}.$ It is obvious that ${\mathcal{R}}^*$ is not cyclic and ${\mathcal{R}}={\mathcal{R}}^*\cup M$.

The \emph{Trace function,} denoted by $Tr()$, of $a+bu$ over $\mathcal{R}$ is defined as
$$Tr(a+bu)=tr(a)+tr(b)u,$$ for $a,b\in \F_{p^m},$
where $tr()$ denotes the trace function from $\F_{p^m}$ onto $\F_p$. Then $R$-linearity of $Tr()$ follows from the $\F_p$-linearity of $tr()$.


%
%

\subsection{The weight formula of $C_D$}
Let $N$ be a positive integer such that $N|p^m-1$. Let $N_1=lcm(N, \frac{p^m-1}{p-1})$
 and $N_2=gcd(N, \frac{p^m-1}{p-1})$. Now, we recall a lemma about abelian groups.\ \

\noindent{\bf  Lemma 2.1}\cite{J} Let $H$ and $K$ be two subgroups of a finite
abelian group $G$. Then $h_1K = h_2K$ if and only if $h_1(H \cap K)=h_2(H\cap K)$ for $h_1, h_2 \in H$. Moreover, there is an
isomorphism: $HK/K\cong H/(H \cap K)$ and $[HK : K] = [H :
(H\cap K)],$ where $HK =\{hk:h\in H,k\in K\}$.\vspace*{0.2cm}

Next, we use some notations as follows. Let $\alpha$ be a fixed primitive element of $\F_{p^m}$. Let $\phi,\chi$ denote the canonical additive characters of $\F_p$ and $\F_{p^m}$, respectively. Let $\lambda,\psi$ denote the multiplicative characters of $\F_p$ and $\F_{p^m}$, respectively. The Gauss sums over $\F_p$ and $\F_{p^m}$ are defined respectively by
$$G(\lambda,\phi)=\sum_{x\in \F_p^*}\lambda(x)\phi(x),~~~G(\psi,\chi)=\sum_{x\in \F_{p^m}^*}\psi(x)\chi(x).$$
Define $C_i^{N}=\alpha^i\langle\alpha^N\rangle$ for $i=0,1,\dots,N-1,$ where $\langle\alpha^N\rangle$ denotes the subgroup of $\F_{p^m}^*=\langle\alpha\rangle$ generated by $\langle\alpha^N\rangle$. In fact, the cosets $C_i^N$ are called the cyclotomic classes of order $N$ in $\F_{p^m}$. Note
that $\F_{p}^*=\langle\alpha^{\frac{p^m-1}{p-1}}\rangle$. Denote $H =C_0^N$ and $K = \F_{p}^*$. Then
we have $H\cap K = C_0^{N_1}$ and $HK = C_0^{N_2}$. Let
$n = [H : (H\cap K)] = |H|/|H \cap K| = N_1/N.$ There is a coset decomposition of $H$ as follows:
$$H=\bigcup_{j=1}^nh_j(H\cap K ),$$
where $h_j=\alpha^{N(j-1)},j=1,2,\dots,n$. By using Lemma 2.1,
we have the coset decomposition of $HK$:
\begin{equation}\label{1}
 HK=\bigcup_{j=1}^nh_jK.
\end{equation}
Define $d_j=h_j=\alpha^{N(j-1)}$ for $j=1,2,\dots,n$, where $n=\frac{N_1}{N}.$ Then $$D=\{d_j=\alpha^{N(j-1)}:j=1,2,\dots,n\}\subseteq C_0^N\subseteq \F_{p^m}.$$
By Equation (1), $d_1, d_2, \dots, d_n$ form a complete set of coset representatives
of the factor group $C_0^{N_2}/\F_p^*$.

Write $c_b=(tr(bd_1),tr(bd_2),\dots,tr(bd_n))$. Denoted by $w_H(c_b)$ the Hamming weight of $c_b$. The code $C_D$ is then defined as $C_D=\{c_b:b\in \F_{p^m}\}$. Note that $C_D$ is punctured from the linear code defined in \cite{SWLP} up to coordinate permutations.
And $w_H(c_b) =
n-N(b)$, where $$N(b)=| \{1\leq j\leq n:tr(bd_j)=0 \}|.$$
By the basic facts of additive characters and Formula (1), we have
\begin{eqnarray*}
  pN(b) &=& \sum_{j=1}^n\sum_{y\in \F_p}\phi(ytr(bd_j))
   =n+\sum_{j=1}^n\sum_{y\in \F_p^*}\chi(ybd_j)\\
    &=&  n+\sum_{x\in C_0^{N_2}}\chi(bx)
   = n+\frac{1}{N_2}\sum_{x\in \F_{p^m}^*}\chi(bx^{N_2}) \\
   &=& n+\frac{1}{N_2(p^m-1)}\sum_{x\in \F_{p^m}^*}\sum_{\psi\in \hat{\F}_{p^m}^*}G(\bar{\psi},\chi)\psi(bx^{N_2})\\
   &=& n+\frac{1}{N_2(p^m-1)}\sum_{\psi\in \hat{\F}_{p^m}^*}G(\bar{\psi},\chi)\psi(b)\sum_{x\in \F_{p^m}^*}\psi(x^{N_2}).
\end{eqnarray*}
By the orthogonally property of multiplicative characters \cite{LN}, we know
\begin{equation*}
  \sum_{x\in \F_{p^m}^*}\psi(x^{N_2})=\begin{cases}
p^m-1,~~~\mathrm{if} ~\psi^{N_2}=\psi_0~(\mathrm{trivial}~\mathrm{character}~\mathrm{of}~\F_{p^m}),\\
  0,~~~~~~~~~~\mathrm{oterwise}.
   \end{cases}
\end{equation*}
Hence,
\begin{equation}\label{3}
  pN(b)=n+\frac{1}{N_2}\sum_{j=0}^{N_2-1}G(\bar{\varphi^j},\chi)\varphi^j(b),
\end{equation}
where $\varphi$ is a multiplicative character of order $N_2$ in $\hat{\F}_{p^m}^*$. Here, $\hat{\F}_{p^m}^*$ denotes multiplicative character group. Then
the weight formula of $C_D$ follows.

Take $q=p$ in Theorem 4.1 of \cite{HY}, then it is easy to obtain the following theorem.\\
\noindent{\bf Theorem 2.2} Assume $m$ to be even, and $N_2=gcd(N,\frac{p^m-1}{p-1})>2$ with $N| (p^m-1)$. Assume that there exists a positive integer $k$ such that $p^k\equiv -1~(\mathrm{mod}~ N_2)$. Denote $t=\frac{m}{2k} $.
\begin{itemize}
 \item[(1)] If $N_2$ is even, $p,t,$ and $\frac{p^k+1}{N_2}$ are odd, then the linear
code $C_D$ defined in (2) is a two-weight $[\frac{N_1}{N},m]$ linear code provided that $N_2<p^{\frac{m}{2}}+1$,
with two weights $\varpi_1<\varpi_2$ of values $\varpi_1 =\frac{p^m-(N_2-1)p^{\frac{m}{2}}}{pN_2}$
 and $\varpi_2 =\frac{p^m+p^{\frac{m}{2}}}{pN_2}$, with respective frequencies $f_1$ and $f_2$ given by
 $f_1=\frac{p^m-1}{N_2}$ and  $f_2=\frac{(N_2-1)(p^m-1)}{N_2}$.
 \item[(2)] In all other cases, the linear code $C_D$ defined in (2)
is a two-weight $[\frac{N_1}{N},m]$ linear code provided that $p^{\frac{m}{2}}+(-1)^t(N_2-1)>0$, with two weights $\varpi_1, \varpi_2$ of values $\varpi_1 =\frac{p^m+(-1)^t(N_2-1)p^{\frac{m}{2}}}{pN_2}$
 and $\varpi_2 =\frac{p^m-(-1)^tp^{\frac{m}{2}}}{pN_2}$, with respective frequencies $f_1$ and $f_2$ given by
 $f_1=\frac{p^m-1}{N_2}$ and  $f_2=\frac{(N_2-1)(p^m-1)}{N_2}$.
 \end{itemize}

For the rest of this paper, for convenience, we adopt the following notations unless otherwise stated
in this paper. Setting $L=\{a+bu:a\in D,b\in \F_{p^m}\}\subseteq \mathcal{R}^*$. So $|L|=np^m.$

\subsection{Codes and Gray map}
  A {\bf linear code} $C$ over $R$ of length $n$ is an $R$-submodule of $R^n$. For $x=(x_1,x_2,\dots,x_n),y=(y_1,y_2,\dots,y_n)\in R^n$, their standard inner product
  is defined by $\langle x,y\rangle=\sum_{i=1}^nx_iy_i$, where the operation is performed in $R$.
  Let $C$ be a linear code over $R$. The {\bf dual code} $C^\perp$ of $C$ consists of all vectors of $R^n$
which are orthogonal to every codeword in $C$, that is, $C^\perp=\{y\in R^n|\langle x,y\rangle =0, \forall x\in C\}.$

The definition of the Gray map $\Phi$ from $R$ to $\F_p^2$ is defined as $\Phi(a+bu )=(b,a+b ),$ where $a,b \in \F_p.$ This map $\Phi$ can be extended to $R^n$ in an obvious way. As observed in \cite{SWLP}, $\Phi$ is a distance preserving isometry from $(R^n,d_L)$ to $(\F_p^{2n},d_H)$, where $d_L$ and $d_H$ denote the Lee and Hamming distance in $R^n$ and $\F_p^{2n}$, respectively. What is more, if $C$ is a linear code over $R$ with parameters $(n,p^k,d)$, then $\Phi(C)$ is a linear code over $\F_p$ with parameters $[2n,k,d]$.

 Given a finite abelian group $G,$ a code over $R$ is said to be {\bf abelian} if it is an ideal of the group ring $R[G].$ In other words, the coordinates of $C$ are indexed by elements of $G$, and $G$ acts regularly on this set. In the special case when $G$ is cyclic, the code is a cyclic code in the usual sense \cite{MS}.


\section{Symmetry}
First, for $a\in \mathcal{R}$ define the vector $Ev(a)$ by the evaluation map $Ev(a)=(Tr(ax))_{x\in L }.$ Define the code $C(m,p,N)$ by the formula $C(m,p,N)=\{Ev(a) :a\in \mathcal{R}\}$. Thus $C(m,p,N)$ is a code of length $|L|=\frac{N_1}{N}p^m$, over $R$. The following result is a simple generalization of Proposition 3.1 in \cite{SWLP}.\\
\textbf{Proposition 3.1} The subgroup $L$ of $\mathcal{R}^*$ acts regularly on the coordinates of $C(m,p,N).$

The code $C(m,p,N)$ is thus an {\em abelian code} with respect to the group $L.$ In other words, it is an ideal of the group ring $R[L].$ As observed in the previous section $L$ is a not cyclic group, hence $C(m,p,N)$ may be not cyclic.

%
%

\section{The Lee Weight of $C(m,p,N)$}
For convenience, we adopt some notations unless otherwise stated
in this paper. Let $\omega=\exp(\frac{2\pi i}{p})$ and $s=2|L|=2\frac{N_1}{N}p^{m}$. If $y=(y_1,y_2,\dots,y_s)\in \mathbb{F}_p^s,$ let $$\Theta(y)=\sum_{j=1}^s\omega^{y_j}.$$
For simplicity, we let $\theta(a)=\Theta(\Phi(Ev(a))).$ By linearity of the Gray map, and of the evaluation map, we see that $\theta(\tau a)=\Theta(\Phi(Ev(\tau a))),$ for any $\tau\in \F_p^*.$

In order to determine the Lee weight of the codewords of $C(m,p,N)$, we first recall the following three lemmas.\\
\textbf{Lemma 4.1}~\cite{SWLP}\label{5.1} For all $y=(y_1,y_2,\dots,y_s)\in \mathbb{F}_p^s,$ we have
$$\sum_{\tau=1}^{p-1}\Theta(\tau y)=(p-1)s-pw_H(y).$$
\textbf{Lemma 4.2}~\cite{SWLP} Let $\Re(\Delta)$ denote the real part of the complex number $\Delta.$ If $p\equiv 3 \pmod{4},$ then $$\sum_{s=1}^{p-1}\theta(sa)=(p-1)\Re(\theta(a)).$$
\textbf{Lemma 4.3}~\cite{MS} If $z \in \mathbb{F}_{p^m}^*,$ then $$\sum\limits_{x\in \mathbb{F}_{p^m}}\omega^{tr(z x)}=0.$$

According to Lemma 4.1, for $Ev(a)\in C(m,p,N)$, by definition of the Gray map, we have
\begin{eqnarray}
 w_L(Ev(a) ) &=& \frac{(p-1)s- \sum\limits_{\tau=1}^{p-1}\Theta(\tau\Phi(Ev(a)))}{p} \nonumber \\
 &=& \frac{(p-1)s-\sum\limits_{\tau=1}^{p-1} \theta(\tau a)}{p} .
\end{eqnarray}
 We are now ready to discuss the Lee weight of the
codewords of the abelian codes introduced above.

\noindent{\bf Theorem 4.4} Let $N_2=1$. If one of the following holds:
$$\mathrm{(i)}~m~\mathrm{is}~\mathrm{even},~~~\mathrm{(ii)}~p\equiv3~(\mathrm{mod}~4)~\mathrm{when}~ m~ \mathrm{is}~ \mathrm{odd},$$
then the set $C(m,p,N)$ is a two-Lee-weight code and its weight distribution is given in Table I.
\begin{center}$\mathrm{Table~ I. }~~~\mathrm{weight~ distribution~ of}~ C(m,p,N) $\\
\begin{tabular}{cccc||cc}
\hline
  Weight&&   & & Frequency  \\
  \hline

  0        & &   & & 1\\
  $2p^{2m-1}$        & &   &              &$p^m-1$\\
  $2p^{2m-1}-2p^{m-1} $  &    & &       &$p^m(p^m-1)$ \\
  \hline
\end{tabular}
\end{center}

\begin{proof}
 Assume $\F_{p^m}^*=\langle\alpha\rangle,$ we get then $\F_p^*=\langle\alpha^{\frac{p^m-1}{p-1}}\rangle$. Let $x=t+t'u$, where $t\in D,t'\in \F_{p^m}$. If $a=0$, then $Ev(a)=(\underbrace{0,0,\dots,0}\limits_{|L|})$. So $w_L(Ev(a))=0$. If $a=\beta u,$ where $\beta\in \F_{p^m}^{*}$. Then $ax=t\beta u$ and $$Tr(ax)=tr(t\beta)u.$$
Taking Gray map yields $$\Phi(Ev(a))=(tr(t\beta),tr(t\beta))_{t,t'}.$$
Since $\Phi$ is a distance preserving isometry from $(R^n,d_L)$ to $(\F_p^{2n},d_H)$,
we have $w_L(Ev(a))=w_H(\Phi(Ev(a)))=2p^m(n-N(\beta))$, where $N(\beta)=|\{1\leq j\leq n: tr(d_j\beta)=0 \}|.$ Note that $N_2=1$. By Formula (3), we know $pN(\beta)=\frac{p^m-p}{p-1}$, which implies $w_L(Ev(a))=2p^{2m-1}.$

If $a=r_0+r_1u\in \mathcal{R}^*$, where $r_0\in\F_{p^m}^*,r_1\in \F_{p^m}$, then $Tr(ax)=tr(r_0t)+tr(r_0t'+r_1t)u$. Taking Gray map yields $$\Phi(Ev(a))=(tr(r_0t'+r_1t),tr(r_0t'+r_1t+r_0t))_{t,t'}.$$
Taking character sums and using Lemma 4.3  yields
$$\theta(a)=\sum_{t\in D}\sum_{t'\in \F_{p^m}}\omega^{tr(r_0t'+r_1t)}+\sum_{t\in D}\sum_{t'\in \F_{p^m}}\omega^{tr(r_0t'+r_1t+r_0t)}=0.$$
 Suppose $m$ is even. Note that $2|\frac{p^m-1}{p-1}$. Thus we claim that any $s\in \F_p^*$ is a square in $\F_{p^m}$, which implies $ \theta(sa)= \theta(a)$.  By Formula (3), we have $w_L(Ev(a))=\frac{(p-1)s}{p}$.
 Suppose that $m$ is odd, and that $p\equiv 3 \pmod{4}$. By using Lemma 4.2, $\sum_{\tau=1}^{p-1} \theta(\tau a)=0$. Further, we have $w_L(Ev(a))=\frac{(p-1)s}{p}$.
\end{proof}\vspace*{0.2cm}
Now, we investigate the dual distance of $C(m,p,N)$ in Theorem 4.4. The proof of the following theorem is similar to that in \cite{SWLP}, so we omit it here. \\
\noindent{\bf Theorem 4.5} Let $N_2=1$. If one of the following holds:
$$\mathrm{(i)}~m~\mathrm{is}~\mathrm{ even},~~~\mathrm{(ii)}~p\equiv3~(\mathrm{mod}~4)~\mathrm{when}~ m~ \mathrm{is}~ \mathrm{odd},$$
 then the dual Lee distance $d'$ of $C(m,p,N)$ is 2. \vspace*{0.2cm}\ \

According to Theorem 4.4, we have constructed a $p$-ary code of length $s=2\frac{p^m-1}{p-1}p^{m},$ dimension $2m,$ with two weights $\omega_1<\omega_2$  of values
$$\omega_1=2p^{2m-1}-2p^{m-1},~~~\omega_2=2p^{2m-1},$$
with respective frequencies $f_1,f_2$ given by
$$f_1=p^{m}(p^m-1),~~~f_2=p^{m}-1.$$

\noindent{\bf Remark 4.6} Despite many similarities, the lengths and weights of the above two-weight codes over $\F_p$ are different from these in the literature \cite{CK}. If $m$ is odd and $p=3$, then the codes in this paper are the same as the linear code in Theorem 5.4 of \cite{SWLP}. If $p\neq3$ or $m$ is even, then their parameters are different.\vspace*{0.2cm}

Next, we study their optimality.\\
\noindent{\bf Theorem 4.7} Suppose $N_2=1$.
If one of the following holds:
$$\mathrm{(i)}~m~\mathrm{is}~\mathrm{ even},~~~\mathrm{(ii)}~p\equiv3~(\mathrm{mod}~4)~\mathrm{when}~ m~ \mathrm{is}~ \mathrm{odd},$$
 then the code $\Phi(C(m,p,N))$ meets the Griesmer bound with equality.
\begin{proof}
Recall the $p$-ary version of the Griesmer bound \cite{G}. If $[N,K,d]$ are the parameters of a linear $p$-ary code, then $$\sum_{j=0}^{K-1}\Big\lceil\frac{d}{p^j}\Big\rceil\leq N.$$
In our situation $N=\frac{2p^{2m}-2p^m}{p-1},K =2m,d =2p^{2m-1}-2p^{m-1}.$
The ceiling
function takes two values depending on the position of $j.$
\begin{itemize}
 \item $j\le m-1 \Rightarrow \lceil \frac{d}{p^j} \rceil =2p^{m-1-j}(p^m-1),$
 \item $j> m-1 \Rightarrow \lceil \frac{d}{p^j} \rceil =2p^{2m-j-1}.$
\end{itemize}
Thus,
\begin{eqnarray*}
  \sum_{j=0}^{K-1}\Big\lceil\frac{d}{p^j}\Big\rceil &=&   \sum_{j=0}^{m-1}\Big\lceil\frac{d}{p^j}\Big\rceil +\sum_{j=m}^{2m-1}\Big\lceil\frac{d}{p^j}\Big\rceil \\
   &=&  \sum_{j=0}^{m-1}(2p^{m-1-j}(p^m-1))+\sum_{j=m}^{2m-1}2p^{2m-j-1} \\
   &=& \frac{2p^{2m}-2p^m}{p-1}=N .
\end{eqnarray*}
The proof is completed.
\end{proof}\vspace*{0.2cm}

\noindent{\bf Example 4.8}
 Let $p=7,m=2$ and $N=3.$ Note that $gcd(N,\frac{p^m-1}{p-1})=gcd(3,8)=1=N_2.$ By Theorem 4.4, we obtain a two-weight linear code over $\F_7$ with parameters $[784,4,672]$. Further, Theorem 4.7 tell us that the code is optimal.

\noindent{\bf Theorem 4.9} Suppose $1<N_2<\sqrt{p^m}+1$.
If one of the following holds:
$$\mathrm{(i)}~m~\mathrm{is}~\mathrm{ even},~~~\mathrm{(ii)}~p\equiv3~(\mathrm{mod}~4)~\mathrm{when}~ m~ \mathrm{is}~ \mathrm{odd},$$
 then $C(m,p,N)$ is a  $(|L|,p^{2m'},d_L)$ linear code over $R$ which has at most $N_2+1$ nonzero Lee weights, where $m'\leq m$ and $$\frac{2p^{m-1}[p^m-(N_2-1)p^{\frac{m}{2}}]}{N_2}\leq d_L(C(m,p,N))\leq\frac{2p^{m-1}(p^m-1)}{N_2}.$$
\begin{proof} Let $x=t+t'u$, where $t\in D, t'\in \F_{p^m}$. For a nonzero codeword $Ev(a)\in C(m,p,N)$, if $a=\beta u\in M\backslash\{0\}$, where $\beta\in \F_{p^m}^*,$
then $\Phi(Ev(a))=(tr(t\beta), tr(t\beta) )_{t,t'}$ and $w_L(Ev(a))=w_H(\Phi(Ev(a)))=2p^m(n-N(\beta))$, where $N(\beta))=|\{1\leq j\leq n: tr(d_j\beta)=0 \}|.$ By Formula (2), we know
\begin{eqnarray*}
  n-N(\beta) &=& n-\frac{n}{p}-\frac{\sum_{j=0}^{N_2-1}G(\bar{\varphi^j},\chi)\varphi^j(b)  }{pN_2} \\
  &=& \frac{n(p-1)}{p}-\frac{-1+\sum_{j=1}^{N_2-1}G(\bar{\varphi^j},\chi)\varphi^j(b)  }{pN_2} \\
   &=& \frac{p^m}{pN_2}-\frac{\sum_{j=1}^{N_2-1}G(\bar{\varphi^j},\chi)\varphi^j(b)  }{pN_2}.
\end{eqnarray*}
Note that $$\Big| \sum_{j=1}^{N_2-1}G(\bar{\varphi^j},\chi)\varphi^j(b) \Big|\leq(N_2-1)p^{\frac{m}{2}}.$$
So $$2p^{m-1}\frac{p^m-(N_2-1)p^{\frac{m}{2}}}{N_2}\leq w_L(Ev(a))\leq  2p^{m-1}\frac{p^m+(N_2-1)p^{\frac{m}{2}}}{N_2},~~~~~~~~~~~~~~a\in M\backslash\{0\},$$
due to $N_2<p^{\frac{m}{2}}+1$. Note that $n-N(\beta)$ is exactly the Hamming weight of the codeword of the cyclic code $C_D$. However, $C_D$ has at most $N_2$ nonzero weights.

Set $a=r_0+r_1u\in \mathcal{R}^*,$ where $r_0\in \F_{p^m}^*,r_1\in \F_{p^m}$. The case is like in the proof of Theorem 4.4. Thus, we can get $w_L(Ev(a))=\frac{(p-1)s}{p}=2p^{m-1}\frac{p^m-1}{N_2}.$ Note that $2p^{m-1}\frac{p^m-1}{N_2}<2p^{m-1}\frac{p^m+(N_2-1)p^{\frac{m}{2}}}{N_2}$. Hence,
$$\frac{2p^{m-1}[p^m-(N_2-1)p^{\frac{m}{2}}]}{N_2}\leq d_L(C(m,p,N))\leq\frac{2p^{m-1}(p^m-1)}{N_2}.$$
 Thus, if $a$ runs through $\mathcal{R}$, the code $C(m,p,N)$ has at most $N_2+1$ different Lee weights. The result follows.
\end{proof}

Next, we investigate the weight distribution of $C(m,p,N)$ if there exists a positive integer $k$ such that $p^k\equiv -1~(\mathrm{mod}~ N_2)$.\\
\noindent{\bf Theorem 4.10} Assume $m$ even, and $N_2=gcd(N,\frac{p^m-1}{p-1})>2$ with $N| (p^m-1)$. Assume that there exists a positive integer $k$ such that $p^k\equiv -1~(\mathrm{mod}~ N_2)$. Denote $t=\frac{m}{2k} $.
\begin{itemize}
 \item[(1)] If $N_2$ is even, $p,t,$ and $\frac{p^k+1}{N_2}$ are odd, then the linear
code $C(m,p,N)$ is a three-Lee-weight linear code provided that $N_2<p^{\frac{m}{2}}+1$ and its weight is given in Table II.
\begin{center}$\mathrm{Table~ II.} ~~~\mathrm{weight~ distribution~ of}~ C(m,p,N) $\\
\begin{tabular}{cccc||cc}
\hline
  Weight&&   & & Frequency  \\
  \hline

  0        & &   & & 1\\
  $\frac{2p^{m-1}[p^m-(N_2-1)p^{\frac{m}{2}}]}{N_2}$        & &   &              &$\frac{p^m-1}{N_2}$\\
  $\frac{2p^{m-1}(p^m-1)}{N_2}$  &    & &       &$p^m(p^m-1)$ \\
   $\frac{2p^{m-1}[p^m+p^{\frac{m}{2}}]}{N_2}$        & &   &              &$\frac{(N_2-1)(p^m-1)}{N_2}$\\
  \hline
\end{tabular}
\end{center}
 \item[(2)] In all other cases, the linear code $C(m,p,N)$
is a three-Lee-weight linear code provided that $p^{\frac{m}{2}}+(-1)^t(N_2-1)>0$ and its weight is given in Table III.
\begin{center}$\mathrm{Table~ III}. ~~~\mathrm{weight~ distribution~ of}~ C(m,p,N) $\\
\begin{tabular}{cccc||cc}
\hline
  Weight&&   & & Frequency  \\
  \hline

  0        & &   & & 1\\
  $\frac{2p^{m-1}[p^m+(-1)^t(N_2-1)p^{\frac{m}{2}}]}{N_2}$        & &   &              &$\frac{p^m-1}{N_2}$\\
  $\frac{2p^{m-1}(p^m-1)}{N_2}$  &    & &       &$p^m(p^m-1)$ \\
   $\frac{2p^{m-1}[p^m-(-1)^tp^{\frac{m}{2}}]}{N_2}$        & &   &              &$\frac{(N_2-1)(p^m-1)}{N_2}$\\
  \hline
\end{tabular}
\end{center}
\end{itemize}
\begin{proof}By a similar approach in the proof of Theorem 4.4, and applied to the correlation Theorem 2.2, the result follows.
\end{proof}

\noindent{\bf Remark 4.11}  By using Theorem 4.10, we obtain a family of $p$-ary three-weight codes $\Phi(C(m,p,N))$.

\section{Application to secret sharing schemes}
Secret sharing is an important topic of cryptography, which has been studied for over thirty years. In this section, we will study the secret sharing schemes based on linear codes studied in this paper.

\subsection{The access structure of the secret sharing schemes }

If a group of participants can recover the secret by combining
their shares, then any group of participants containing this
group can also recover the secret. A group of participants is referred to
as a minimal access set if they can recover the secret with
their shares, but any of its proper subgroups cannot do so.
Here a proper subgroup has fewer members than this group.
Based on these facts, we are only interested in the set of all
minimal access sets. Thus the concept of a minimal codeword of a linear code $C$ is introduced. A \emph{minimal codeword} of a linear code $C$ is a nonzero codeword that does not cover any other nonzero codeword. We say that a vector $x$ covers a vector $y$ if $s(x)$ contains $s(y)$, where $s(y)$, the support $s(y)$ of a vector $y\in \F_q^s$, is defined as the set of indices where it is nonzero. Although determining the minimal codewords
of a given linear code is a difficult problem in general, there is a numerical condition derived in \cite{AB}, bearing on the weights of the code, that is easy to check. One of conditions about determining the minimal codewords of a given linear code described by the following lemma \cite{AB}.\\
{\bf Lemma 5.1 }(Ashikmin-Barg) Denote by $w_0$ and $w_{\infty}$ the minimum and maximum nonzero weights of a $q$-ary code $C$, respectively. If
$$\frac{w_0}{w_{\infty}}>\frac{q-1}{q},$$ then every nonzero codeword of $C$ is minimal. \

We can infer from there the support structure for the codes of this paper.\\
{\bf Theorem 5.2} If one of the following holds:
$$\mathrm{(i)}~m~\mathrm{is}~\mathrm{ even},~~~\mathrm{(ii)}~p\equiv3~(\mathrm{mod}~4)~\mathrm{when}~ m~ \mathrm{is}~ \mathrm{odd},$$
 then all the nonzero codewords of $\Phi(C(m,p,N))$, for $N_2=1$, are minimal.
\begin{proof}
 By the preceding Lemma 5.1 with $w_0=\omega_1,$ and $w_{\infty}=\omega_2.$ Rewriting the inequality of Lemma 5.1 as $p\omega_1>(p-1)\omega_2$. Note that
 \begin{eqnarray*}
 p\omega_1 -(p-1)\omega_2&=&p(2p^{2m-1}-2p^{m-1})-2(p-1)p^{2m-1} \\
    &=& 2p^{2m-1}-2p^{m}>0.
 \end{eqnarray*}
 Hence the proposition follows.
\end{proof}

\hspace*{-0.4cm}{\bf Theorem 5.3} Assume $m$ even, and $2<N_2=gcd(N,\frac{p^m-1}{p-1})<p^{\frac{m}{2}-1}$ with $N| (p^m-1)$. Assume that there exists a positive integer $k$ such that $p^k\equiv -1~(\mathrm{mod}~ N_2)$. Denote $t=\frac{m}{2k}$. If $N_2$ is even, $p,t,$ and $\frac{p^k+1}{N_2}$ are odd, then all the nonzero codewords of $\Phi(C(m,p,N))$ are minimal.
\begin{proof}
 We use Lemma 5.1 with $w_0=\frac{2p^{m-1}[p^m-(N_2-1)p^{\frac{m}{2}}]}{N_2},$ and $w_{\infty}=\frac{2p^{m-1}[p^m+p^{\frac{m}{2}}]}{N_2}.$ Rewriting the inequality of Lemma 5.1 as $pw_0>(p-1)w_\infty,$ and dividing both sides by $\frac{2p^{m-1}}{N_2}$, we obtain
 $$p(p^{m}-(N_2-1)p^{\frac{m}{2}})>(p-1)(p^m+p^{\frac{m}{2}}),$$
 or $ N_2p<p^{\frac{m}{2}}$, which is true for $N_2<p^{\frac{m}{2}-1}.$ So the result follows.
\end{proof}
\subsection{Massey's scheme}
Secret sharing was motivated
by the problem of sharing a secret digital key. In order to keep the secret efficiently and
safely, secret sharing scheme (SSS) was introduced in 1979 by Shamir and Blakley. Since then, many applications of SSS to several different kinds of cryptographic protocols have appeared. The so-called Massey's scheme is a construction of such a scheme where a code $C$ of length $S$ over $\F_p$ gives rise to a SSS. The secret is carried by the first coordinate of a codeword, and the coalitions correspond to supports of codewords in the dual code with a one in that coordinate. And the coalition structure is related to the support
structure of $C.$ It is worth mention that in some special cases, that is, when all nonzero codewords are minimal, it was shown in \cite{DY2} that there is the following alternative, depending on $d'$:
\begin{itemize}
 \item If $d'\ge 3,$ then the SSS is \emph{``democratic''}: every user belongs to the same number of coalitions,
 \item If $d'=2,$  then the SSS is  \emph{``dictatorial''}: some users belong to every coalition.
\end{itemize}
Depending on the application, one or the other situation might be more suitable.
By Theorems  4.4,  4.5 and  5.2, we see that for some values of the parameters, a SSS built on $\Phi(C(m,p,N))$ is dictatorial.
\section{Conclusion}

The contributions of this paper include the construction of linear codes with few weights and the
determination of their weight distribution. It is well
known that the weight distribution problem for linear codes
is in general very hard and it is settled for only a very small
number of classes of codes. In the present work, by using an algebraic method and a Gray map, we construct a family of two-weight linear codes, which are optimal by using Griesmer bound, and three-weight linear codes. These codes are not visibly cyclic. It is worth exploring more general constructions by varying the alphabet of the code, or the defining set of the trace code. Compared with cyclic codes in \cite{CK,DLLZ,DY,HY1},  many codes in this paper have different weight distributions.

\section*{References}

\bibliography{mybibfile}

\end{document}